\documentclass[prl,twocolumn,superscriptaddress]{revtex4-1}
\usepackage{amsmath} 
\usepackage{esint}
\usepackage{graphicx}
\usepackage{color}
\usepackage{bm}

\begin{document}

\title{Extremely Persistent Dense Active Fluids}
\author{Grzegorz Szamel}
\affiliation{Department of Chemistry, Colorado State University,
Fort Collins, Colorado 80523, USA}
\author{Elijah Flenner}
\affiliation{Department of Chemistry, Colorado State University,
Fort Collins, Colorado 80523, USA}

\date{\today}

\begin{abstract}
  We examine the dependence of the dynamics of three-dimensional active fluids on persistence time $\tau_p$ and average self-propulsion force $f$.
  In the large persistence time limit many properties of these fluids become $\tau_p$-independent. These properties include the
  mean squared velocity, the self-intermediate scattering function, the shear-stress correlation function and the low-shear-rate viscosity.
  We find that for a given $f$ in the large $\tau_p$ limit the mean squared displacement is independent of the persistence time for times shorter
  than $\tau_p$ and the long-time self-diffusion coefficient is proportional to the persistence time.
  For a large range of self-propulsion forces the large persistence time limits of many properties depend on $f$ as power laws.  
\end{abstract} 

\maketitle
Particles that use energy from their environment to perform persistent motion, \textit{i.e.} self-propelled or active particles, behave
in surprising and interesting ways \cite{Marchetti2013,Elgeti2015,Bechinger2016}.
Recently, novel intermittent
dynamics was identified in extremely persistent dense homogeneous two-dimensional systems \cite{Mandal2020,Mandal2021,Keta2022,Keta2023}.
It was shown that these systems evolve through sequences of mechanical equilibria in which self-propulsion forces balance interparticle interactions.
%The relaxation occurs on the time scale of the persistence time of the self-propulsion, $\tau_p$, and occurs via elastic and plastic events whose size scales
%with the system size.
Here we examine extremely persistent dense homogeneous three-dimensional active fluids in which 
the interparticle interactions never manage to balance the self-propulsion forces. Many
properties of these fluids become $\tau_p$ independent and scale with the root-mean-squared self-propulsion force $f$ as power laws. 

We recall that the phase space of active systems is much larger than that of passive ones. At a minimum, one has to specify the average
strength of active forces and their persistence time in addition to the set of parameters characterizing
the corresponding passive system. If one 
considers athermal active systems, this results in a three-dimensional control parameter space. Thus,
when comparing 
results of diverse studies, one needs to 
specify the path in the parameter space that one is following.

Early studies of dense homogeneous active systems focused on the glassy dynamics and the active glass transition
\cite{Berthier2013,Ni2013,Berthier2014,Szamel2015,Mandal2016,Flenner2016,Berthier2017,Klongvessa2019,Berthier2019,Janssen2019}.
These studies considered a limited range of persistence times and often examined the behavior of the systems at constant active temperature $T_a$
that characterizes the long-time motion of an isolated active particle. At constant $T_a$, with increasing $\tau_p$ the strength of active forces
decreases and dense active systems typically glassify, see Fig. 2c of % Keta \textit{et al.}
Ref. \cite{Keta2022} for a recent
example. Thus, to investigate the effects of extremely persistent active forces %on flowing systems
it is common to fix their strength %of the active forces
while increasing their persistence time \cite{Mandal2020,Mandal2021,Keta2023}.

Recent simulational studies of dense two-dimensional active systems demonstrated that, for
large persistence times, there is a new phase between fluid and glass phases with intermittent dynamics  \cite{Mandal2020,Mandal2021,Keta2023}.
Importantly, in the large $\tau_p$ 
limit the relaxation happens on the time scale of the persistence time, and the mean-square displacement and
the two-point overlap function exhibit well-defined limits when plotted versus time rescaled by the persistence time \cite{Mandal2021}.
This observation suggests that the dynamics at large persistence times may be studied by assuming that
the active and interparticles forces converge to a force-balanced state for times much less than the persistence time and that
all the rearrangements happen on the time-scale of the persistence time. This approach is termed activity-driven dynamics \cite{Mandal2021,MandalSollich2020}.

A recent study that used the activity-driven dynamics algorithm \cite{Keta2022} discovered very interesting extreme persistence time limit
dynamics of two-dimensional active systems. Displacement distributions were found to be non-Gaussian and to exhibit fat exponential tails.
An intermediate-time plateau in the mean-square displacement was absent. Instead, 
a region %of super-diffusive mean-square displacement 
scaling with time as $t^\beta$ with $\beta \approx 1.6$ for times less than the persistence time was identified.
The complex intermittent dynamics resembled that found in zero-temperature driven amorphous solids, but with some important differences.

Here we focus on extremely persistent three-dimensional dense homogeneous active fluids. Like Keta \textit{et al.} \cite{Keta2022,Keta2023},
we %study active fluids in the large $\tau_p$ limit while keeping
use $\tau_p$ and %root-mean-square active force
$f$ as our control parameters. In the systems we investigated, for large enough
$\tau_p$ the mean square velocity saturates at a non-zero value determined by $f$. The interparticle
forces never manage to balance completely the active foces and the systems relax on the time scale shorter than the persistence time.
Therefore, the infinite $\tau_p$ limits of our systems lie in the un-jammed
phase of the active yielding phase diagram studied (in two dimensions) by Liao and Xu \cite{LiaoXu}. 
Several properties of these systems become $\tau_p$-independent and scale as non-trivial power laws with $f$. In the following we describe the
systems we studied and then present and discuss our observations.

We study a three-dimensional 50:50 binary mixture of spherically symmetric active particles interacting via the Weeks-Chandler-Andersen potential,
%\begin{equation}
%\label{pot}
%V_{\alpha \beta} = 4 \epsilon \left[ \left( \frac{\sigma_{\alpha \beta}}{r} \right)^{12} - \left( \frac{\sigma_{\alpha \beta}}{r}\right)^6\right]$
%\end{equation}
$V_{\alpha \beta} = 4 \epsilon \left[ \left(\sigma_{\alpha \beta}/r\right)^{12} - \left(\sigma_{\alpha \beta}/r\right)^6\right]$
for $r < \varsigma_{\alpha \beta} = 2^{1/6} \sigma_{\alpha \beta}$ and 0 otherwise. Here, $\alpha$, $\beta$ denote the
particles species $A$ or $B$ and $\epsilon$ is the unit of energy. The distance unit is set by $\sigma_{BB} = 1.0$, $\sigma_{AA} = 1.4$, and $\sigma_{AB} = 1.2$.
We study the number density $N/V=0.451$, which corresponds to the volume fraction  $\phi=\pi N [\varsigma_{AA}^3 + \varsigma_{BB}^3]/(12 V)=0.625$.

We use the athermal active Ornstein-Uhlenbeck particle %(AOUP)
model %without thermal noise
%to model our active particles 
\cite{Szamel2014,Maggi2015,Fodor2016}.
%In this model the self-propulsion forces evolve according to the Ornstein-Ulenbeck process.
The equation of motion for the position $\mathbf{r}_n$ of particle $n$ is
\begin{equation}\label{eomr}
\xi_0 \dot{\mathbf{r}}_n = \mathbf{F}_n + \mathbf{f}_n,
\end{equation}
where $\mathbf{F}_n = - \sum_{m \ne n} \nabla_n V(r_{nm})$ and $\mathbf{f}_n$ is the active force. $\xi_0=1$ is the friction coefficient
of an isolated particle and $\xi_0\sigma_{BB}/\epsilon$ sets the unit of time.
The equation of motion for $\mathbf{f}_n$ reads
%the self-propulsion force acting on particle $n$ is
\begin{equation}
\tau_p \dot{\mathbf{f}}_n = -\mathbf{f}_n + \boldsymbol{\zeta}_n,
\end{equation}
where $\tau_p$ is the persistence time of the self-propulsion and $\boldsymbol{\zeta}_n$ is
a Gaussian white noise with zero mean and variance
$\left< \boldsymbol{\zeta}_n(t) \boldsymbol{\zeta}_m(t^\prime)\right>_{\mathrm{noise}} = 2 \xi_0 T_a \mathbf{I} \delta_{nm} \delta(t-t^\prime)$,
where $\left< \ldots \right>_{\mathrm{noise}}$ denotes averaging over the noise distribution, $T_a$ is a single particle effective temperature,
$\mathbf{I}$ is the unit tensor and we set the Boltzmann constant $k_B = 1$. The root-mean square strength of active forces is $f = \sqrt{3 T_a/\tau_p}$. 

\begin{figure}
  \includegraphics[width=0.9\columnwidth]{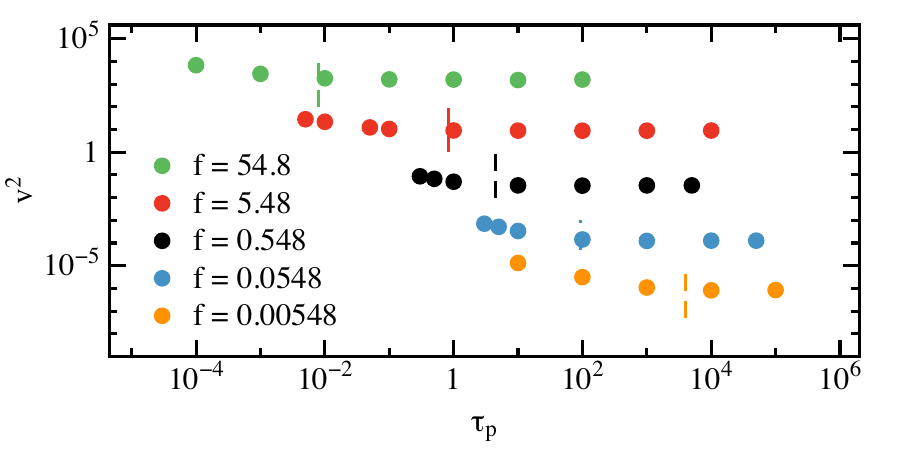}
  \caption{\label{vsq}The persistence time dependence of the mean square velocity $v^2=\left< \dot{\mathbf{r}}^2 \right>$ for fixed active force strength $f$.
    The velocity decreases with $\tau_p$ until $\tau_p(f)$ \cite{method} and then saturates.  The vertical  dashed lines indicate $\tau_p(f)$.}
\end{figure}

We start be examining the persistence time dependence of the mean square velocity,
$v^2 \equiv \left< \sum_n \dot{\mathbf{r}}_n^2 \right> = \left< \sum_n \mathbf{F}_n^2 + \sum_n 2\mathbf{F}_n \cdot \mathbf{f}_n \right> + N f^2$,
%(recall that we set $\xi_0=1$),
which is shown in Fig.~\ref{vsq}. %for $f$ = 0.0548, 0.548, and 5.48.
We observe that with increasing persistence time $v^2$ decreases and then saturates.
For each active force strength $f$ we define a characteristic persistence time $\tau_p(f)$ at which $v^2$ stops changing \cite{method}.
%while with increasing persitence time the particles are able to arrange themselves in such a way that the interparticle forces balance most of the active forces,
The cancellation of the interparticle and %self-propulsion
active forces is never complete, unlike in systems investigated in Ref. \cite{Mandal2021}.
For the range of $f$ that we studied, our systems are never at the bottom of an
effective potential consisting of the potential energy tilted by the terms originating from the active forces \cite{Mandal2021,Keta2023} and do not become arrested on the
time scale of the persistence time. Therefore, in the infinite $\tau_p$ limit our systems fall into the fluid phase of the three-dimensional version of the phase diagram
of Liao and Xu \cite{LiaoXu}.

\begin{figure}
\includegraphics[width=0.9\columnwidth]{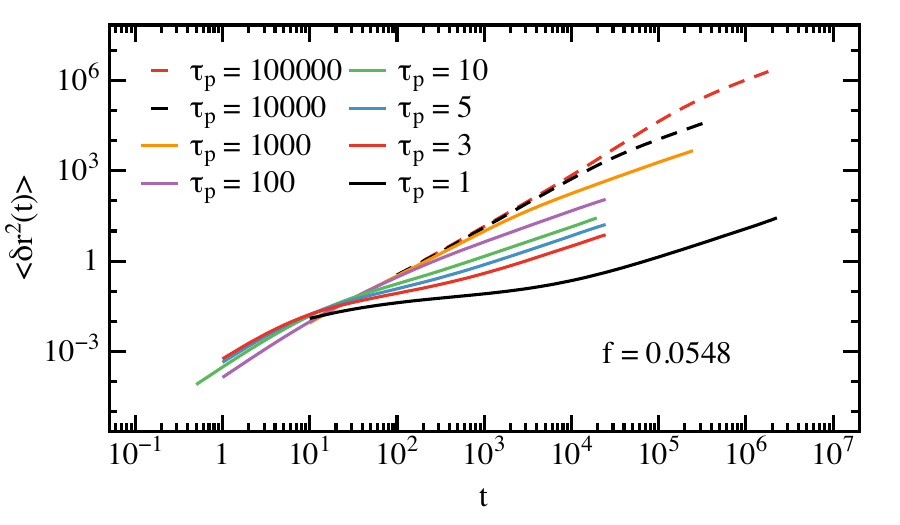}
\includegraphics[width=0.9\columnwidth]{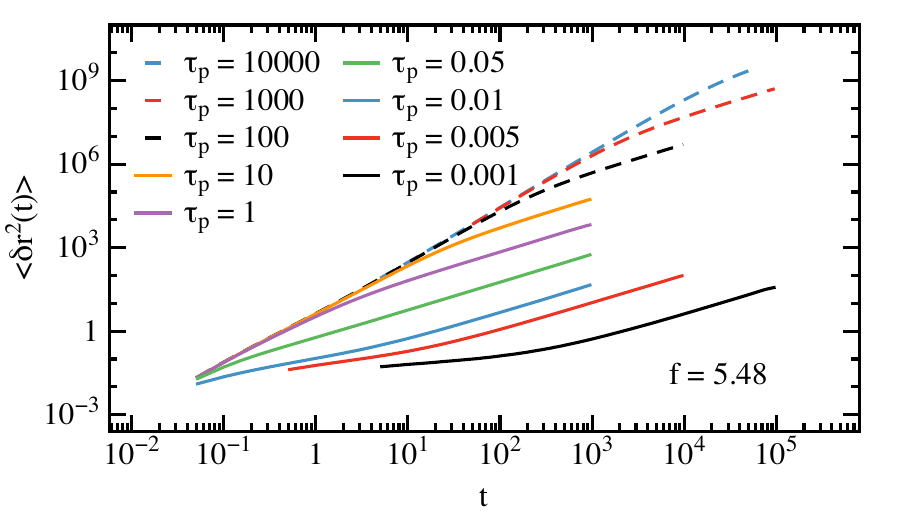}
\caption{\label{msd}The mean-square-displacement $\left< \delta r^2(t) \right>$ at fixed 
  $f$ for several $\tau_p$. For shorter $\tau_p$ the mean squared displacements display a glassy plateau followed by diffusive motion.
  With increasing $\tau_p$ an extended superdiffusive region appears that is analyzed in Fig.~\ref{super}.}
\end{figure}

Next, we examine the persistence time dependence of the mean squared displacement (MSD)
\begin{equation}
\left< \delta r^2(t) \right> = N^{-1} \left< \sum_n [\mathbf{r}_n(t) - \mathbf{r}_n(0)]^2 \right>,
\end{equation}
shown in Fig.~\ref{msd}. %for $f$ = 0.0548, 0.548, and 5.48 and a wide range of $\tau_p$.
At short times  the motion is ballistic and
it is determined by the mean square velocity, $\left< \delta r^2(t) \right> = v^2 t^2$ \cite{Szamel2015}. Thus, the short-times MSD
decreases with increasing $\tau_p$ and becomes constant beyond $\tau_p(f)$. 

\begin{figure}
\includegraphics[width=0.9\columnwidth]{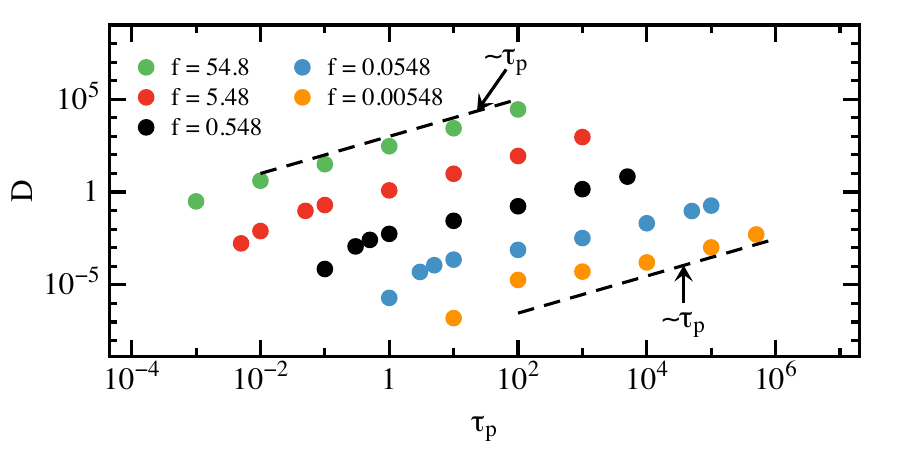}
\caption{\label{diff}The long time diffusion coefficient for fixed active force strength $f$ as a function of persistence time $\tau_p$. 
The diffusion coefficient grows with increasing $\tau_p$ and is proportional to $\tau_p$ in the large $\tau_p$ limit.}
\end{figure}

At long times, the MSD exhibits diffusive behavior. The self-diffusion coefficient, $D=\lim_{t\to\infty} \left< \delta r^2(t) \right>/(6t)$,
is shown in Fig.~\ref{diff}. For a given $f$ it monotonically increases with increasing $\tau_p$.  At
large $\tau_p$ we find that $D \sim \tau_p$, indicated by the dashed lines.

\begin{figure}
\includegraphics[width=0.9\columnwidth]{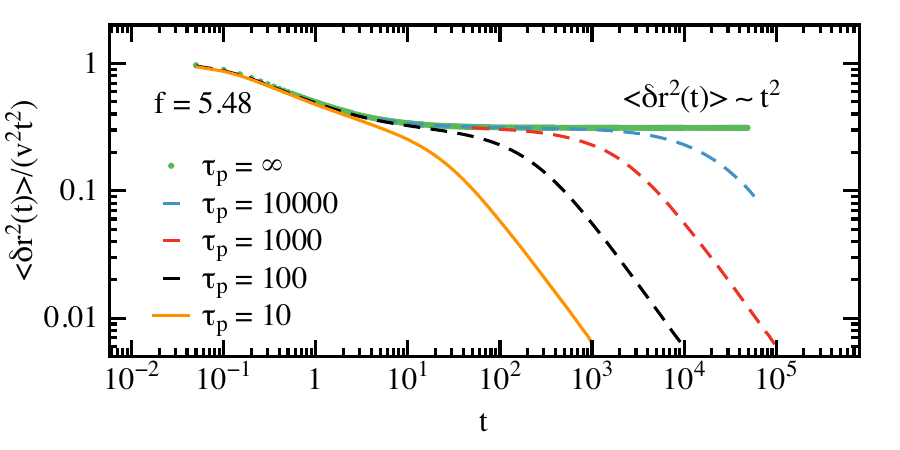}
\caption{\label{super} The mean-square displacement $\left< \delta r^2(t) \right>/(v^2t^2)$ for $f = 5.48$. For
  large $\tau_p$, at short times the motion is ballistic (not shown), and then there is a super-diffusive regime.
  For very large $\tau_p$ the super-diffusive motion speeds up and the resulting behavior is approximately ballistic. 
  For $t>\tau_p$ long-time diffusive motion is observed. In the $\tau_p \rightarrow \infty$ limit the system stays in the second ballistic regime,
  with velocity $v_1$. }
\end{figure}

We found a surprising time-dependence of the MSD between the initial ballistic and the long-time diffusive regimes.
In Fig.~\ref{super} we show the MSD divided by $v^2t^2$ to show this time-dependence more clearly. 
The MSD exhibits a superdiffusive behavior that seems to approach a large $\tau_p$ master curve.
The superdiffusive behavior does not follow a single power law. Instead,
a second, intermediate time ballistic regime appears, with velocity $v_1$. This is in contrast to the finding of Keta \textit{et al.} \cite{Keta2022}
who observed intermediate power law behavior $\left< \delta r^2(t) \right> \sim t^\beta$ with $\beta \approx 1.6$. In the $\tau_p\to\infty$ limit
the systems stays in the second ballistic regime.

Results shown in Fig.~\ref{super} suggest that in the large $\tau_p$ limit the diffusion can be thought of as a random walk consisting of of steps of length
$v_1\tau_p$ taken every $\tau_p$. This picture rationalizes the observed scaling $D\sim\tau_p$.

\begin{figure}
\includegraphics[width=0.9\columnwidth]{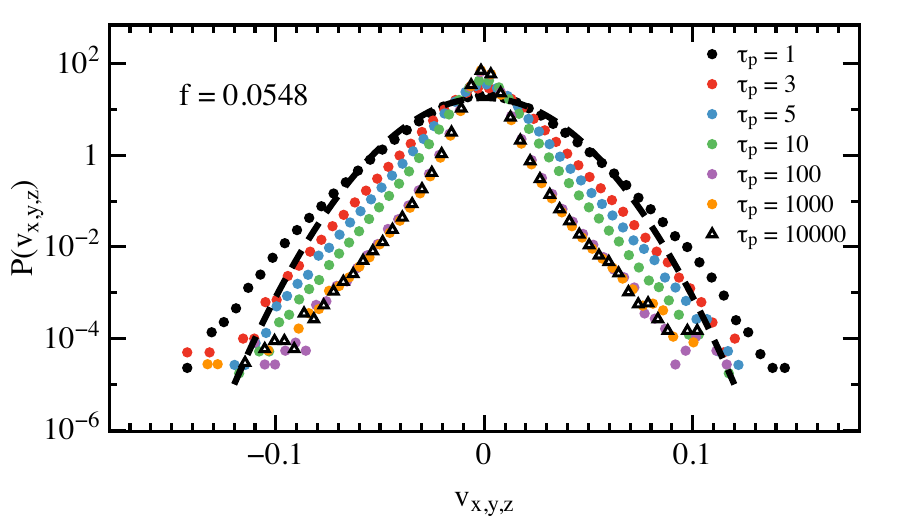}
\caption{\label{veld}The distribution of velocities for $f = 0.0548$ for a broad range of persistence times $\tau_p$.
The distributions are non-Gaussian; the tails become more prominent with increasing $\tau_p$ until $\tau_p(f)$ when the distributions 
overlap. The dashed line represents a Maxwell distribution.
}
\end{figure}

In Fig.~\ref{veld} we show the velocity distributions. As found by Keta \textit{et al.} \cite{Keta2022}, the distributions are strongly non-Gaussian.
Their broad tails become more prominent with increasing $\tau_p$ until $\tau_p(f)$. For $\tau_p > \tau_p(f)$ the distributions overlap.
%The shapes of the velocity distributions are similar to those reported in Ref.~\cite{Keta2022}.

\begin{figure}
\includegraphics[width=0.9\columnwidth]{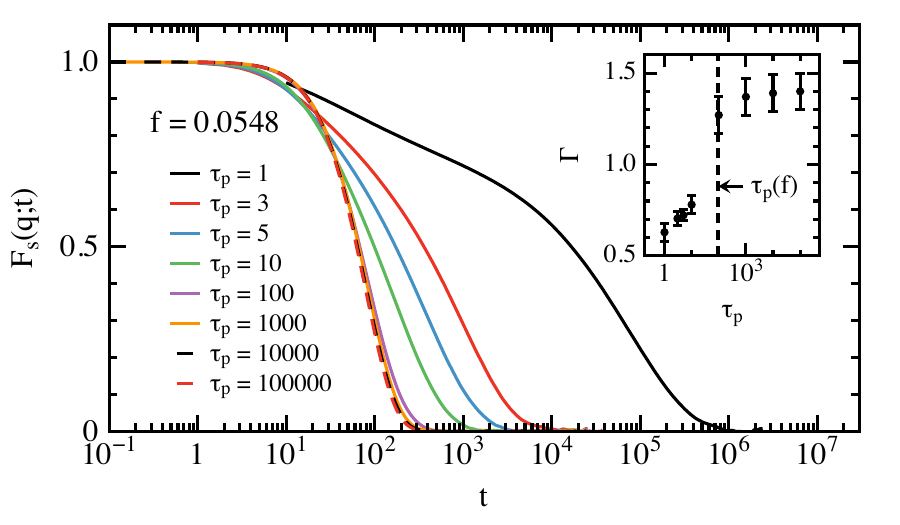}
\caption{\label{fs}The self-intermediate scattering function $F_s(k;t)$ for  $f = 0.0548$ and 
a range of persistence times $\tau_p$. The relaxation time decreases with increasing persistence time until
around $\tau_p(f)$ where $F_s(k;t)$ becomes independent of persistence time. The inset shows the $\tau_p$
dependence of parameter $\Gamma$ of fit to $F_s(k;t) = a e^{-(t/\tau_s)^\Gamma}$ where $a \le 1$.}
\end{figure}

The evolution of the mean square displacement with the persistence time is reflected in the $\tau_p$ dependence of the 
self-intermediate scattering function
\begin{equation}
F_s(k;t) = \frac{1}{N} \left< \sum_n e^{i \mathbf{k} \cdot (\mathbf{r}_n(t) - \mathbf{r}_n(0))} \right>.
\end{equation}
We chose $k = 5.3$, which is approximately equal the first peak of the total static structure factor.
In Fig.~\ref{fs} we show $F_s(k;t)$ for $f = 0.0548$.
%For small $\tau_p$ we observe a plateau region and the stretched exponential decay typically found in supercooled liquids.
With increasing $\tau_p$ the intermediate time glassy plateau disappears and
the decay changes from stretched exponential, to exponential, then to compressed exponential. Shown in the inset to Fig.~\ref{fs}
is the parameter $\Gamma$ obtained from fits to $F_s(k;t) = ae^{-(t/\tau_s)^\Gamma}$ where
we restrict $a \le 1$.  $\Gamma$  increases with increasing $\tau_p$ and reaches a plateau above $\tau_p \approx 94$. % (dashed line in inset).

\begin{figure}
\includegraphics[width=0.9\columnwidth]{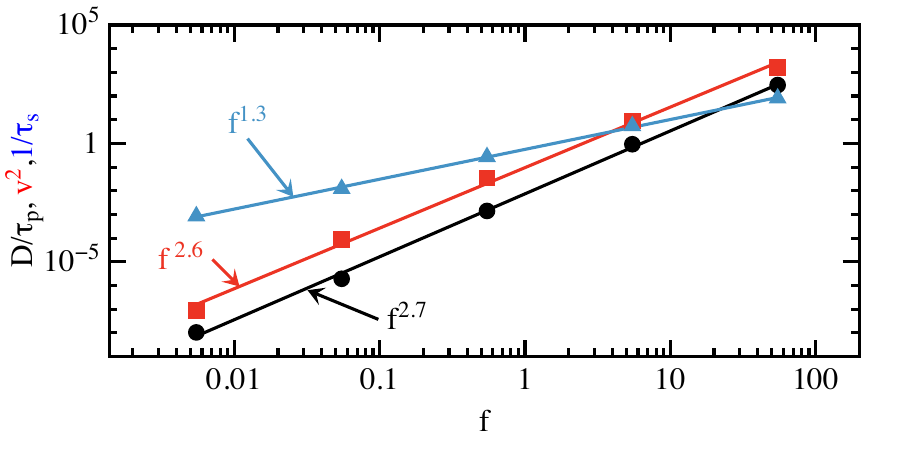}
\caption{\label{power}
  %  (a) The large $\tau_p$ limit of $D/\tau_p$ and $v^2$ as a function of the 
  %  self-propulsion strength $f$. The lines are power law fits where the exponent for $v^2$ and 
  %  $D/\tau_p$ are $2.6 \pm 0.1$ and $2.7 \pm 0.1$, respectively. (b) The large $\tau_p$ limit of 
  %  the relaxation time of $\Sigma_(xy)(t)/\Sigma_{xy}(0)$, $\tau_\eta(f)$, and $F_s(k;t)$, $\tau_\alpha(f)$, 
  %  defined as when these functions
  %  equal $e^{-1}$. The lines are fits to power laws with the exponent $-1.1 \pm 0.1$ for $\tau_\eta(f)$ 
  %  and $-1.3 \pm 0.1$ for $\tau_\alpha(f)$.
  The large $\tau_p$ limit of $v^2$, $D/\tau_p$ and $1/\tau_s$ as a function of the self-propulsion strength $f$.
   The lines are power law fits where the exponent for $v^2$,  $D/\tau_p$ and $1/\tau_s$ are $2.6 \pm 0.1$, $2.7 \pm 0.1$ and $1.3 \pm 0.1$, respectively.
}
\end{figure}

We find that the large persistence time limits of several properties discussed above depend on the strength of the active forces %self-propulsion force
as power laws. In Fig.~\ref{power} we show the large $\tau_p$ limits of $v^2$ (squares), $D/\tau_p$ (circles) and $\tau_s$ (triangles).  
We find that the former two quantities follow a power law with $f$ with statistically the same exponent, $2.6\pm 0.1$ for $v^2$ and $2.5\pm 0.1$ for $D/\tau_p$.
The power law of the relaxation time, $\tau_s\sim f^{-1.3}$ can be related to that of $v^2$; in the large $\tau_p$ limit $F_s$ decays on the time scale on which
a particle moves over its diamater, which scales as $v$.

\begin{figure}
\includegraphics[width=0.9\columnwidth]{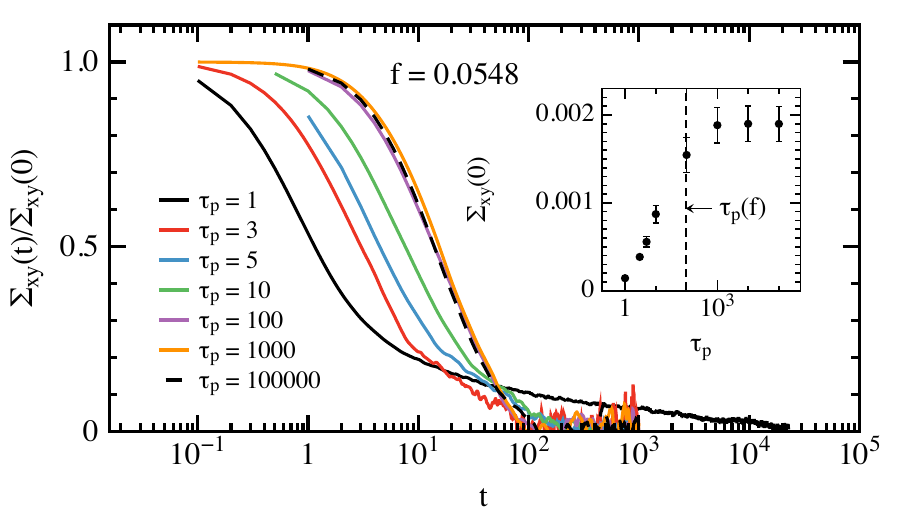}
\caption{\label{sigma}Normalized shear-stress correlation function for $f = 0.0548$
and a large range of persistence times.
The shear stress relaxation time initially increases with $\tau_p$, but becomes constant for $\tau_p > \tau_p(f)$.
Additionally, $\Sigma_{xy}(0)$ grows with $\tau_p$ until approximately $\tau_p(f)$ and then it becomes approximately constant (inset).}
\end{figure}

The above discussed quantities describe the single particle motion in our many-particle systems. To access collective properties of these systems
we investigated the $\tau_p$ dependence of the stress fluctuations and the rheological response. First, we examined the 
shear-stress correlation function $\Sigma_{\alpha \beta}(t) = V^{-1} \left< \sigma_{\alpha \beta}(t) \sigma_{\alpha \beta}(0) \right>$, where 
\begin{equation}
\sigma_{\alpha \beta}(t) = -\frac{1}{2} \sum_n \sum_{m \ne n} \frac{r_{nm}^\alpha r_{nm}^\beta}{r_{nm}} \frac{dV(r_{nm})}{dr_{nm}},
\end{equation}
and $r_{nm}^\alpha$ is the $\alpha$ component of the distance vector between particle $n$ and particle $m$. 
%For $\alpha\neq\beta$ the integral of $\Sigma_{\alpha \beta}/T$
%s the dominant contribution to the viscosity in an equilibrium (non-active) system at low temperature.

In Fig.~\ref{sigma} we report the normalized shear stress correlation function,
$\Sigma_{xy}(t)/\Sigma_{xy}(0)$, for $f = 0.0548$ and a large range of persistence times.
For small $\tau_p$ there is a rapid decay to an emerging plateau followed by a %long,
slow decay to zero. With increasing $\tau_p$, the decay of $\Sigma_{xy}(t)$ becomes more exponential and it 
is exponential above $\tau_p(f)$.  In the inset we show the dependence of the initial value, $\Sigma_{xy}(0)$, on $\tau_p$. 
We see that the initial value first grows with $\tau_p$ and then plateaus.  

\begin{figure}
\includegraphics[width=0.9\columnwidth]{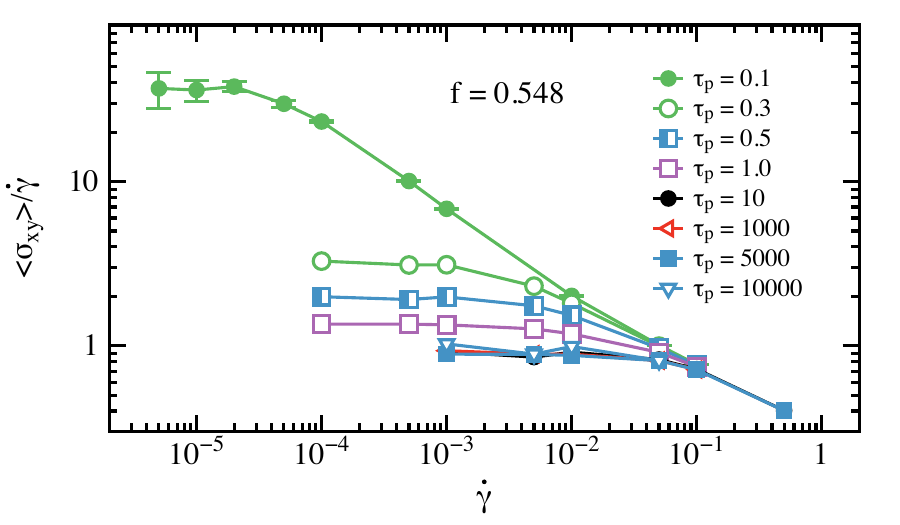}
\caption{\label{flow}Average shear stress $\left<\sigma_{xy}\right>$ divided by
the shear rate $\dot{\gamma}$ at constant $f = 0.548$ for a wide range of
persistence times. The zero-shear-rate viscosity $\eta$ can be obtained from the small shear rate
plateau. There is a decrease in $\eta$ with increasing $\tau_p$ $\eta$ followed by a saturation. }
\end{figure}

To probe the rheological response of our active systems we simulated shear flow by adding to Eq. \eqref{eomr} 
a bulk non-conservative force $\mathbf{F}_n^{\dot{\gamma}} = \xi_0 \dot{\gamma} y_n \mathbf{e}_x$ with
Lees-Edwards boundary conditions \cite{LeesE}. In Fig.~\ref{flow} we show the average shear stress, $\left< \sigma_{xy} \right>/\dot{\gamma}$,
for $f = 0.548$ and a large range of range of $\tau_p$. The flow curves for other strengths of the self-propulsion have the same main features.
%The small $\dot{\gamma}$ behavior is $\tau_p$-dependent, but all the curves collapse at larger $\dot{\gamma}$ for each $f$.
The limiting zero-shear-rate viscosity $\eta$ can be obtained from the small $\dot{\gamma}$ pleateus. 

\begin{figure}
\includegraphics[width=0.9\columnwidth]{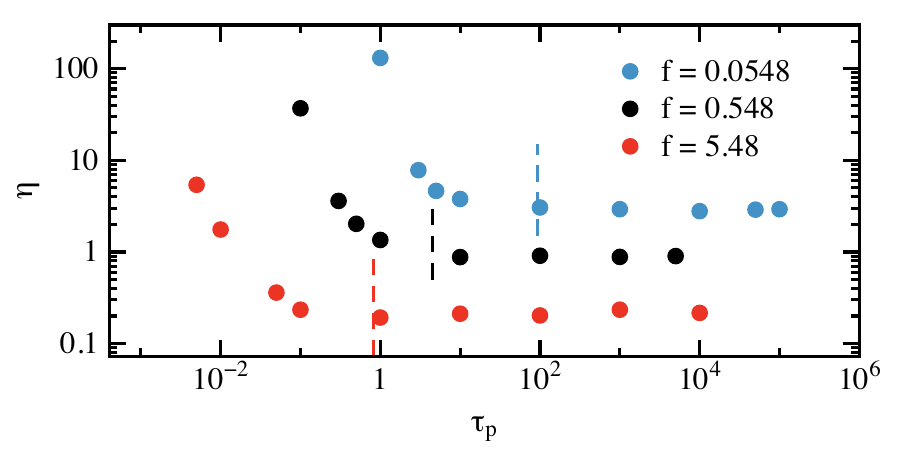}
\caption{\label{eta}The viscosity $\eta$ at fixed average active forces strength versus $\tau_p$. The viscosity 
initially decreases with increasing $\tau_p$ and then becomes constant above $\tau_p(f)$, which is shown as 
vertical dashed lines.}
\end{figure}

In Fig.~\ref{eta} we show the $\tau_p$ dependence of the zero-shear-rate viscosity. %for the three values of $f$ we studied.
We find that $\eta$ initially decreases and reaches a $\tau_p$-independent plateau above $\tau_p(f)$.
%The existence of the plateau is consistent with the unchanging dynamics on
%the time scales corresponding to the motion over one particle diameter.

\begin{figure}
\includegraphics[width=0.9\columnwidth]{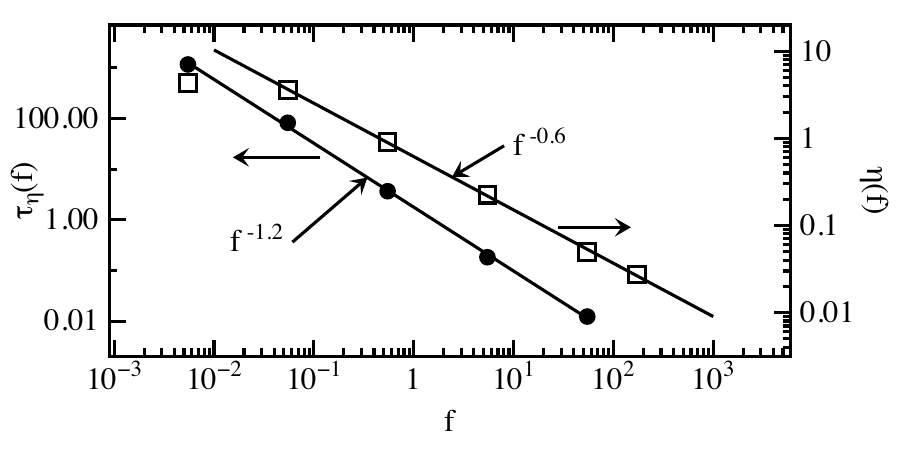}
\caption{\label{taueta} The large $\tau_p$ limit of the relaxation time of $\Sigma_(xy)(t)/\Sigma_{xy}(0)$ (defined as when this function
  equals $e^{-1}$)  and of viscosity $\eta$ as a function of the self-propulsion strength $f$.
  The lines are power law fits where the exponents for $\tau_\eta$ and $\eta$ $-1.2\pm 0.1$ and $-0.60 \pm 0.1$, respectively.}
\end{figure}

Again, we find that large $\tau_p$ limits of collective properties depend on $f$  a power laws. 
In Fig.~\ref{taueta} we show the dependence of the large $\tau_p$ limits of the relaxation time of the normalized stress tensor
autocorrelation function and of the viscosity on the strength of the self-propulsion.

\begin{figure}
\includegraphics[width=0.9\columnwidth]{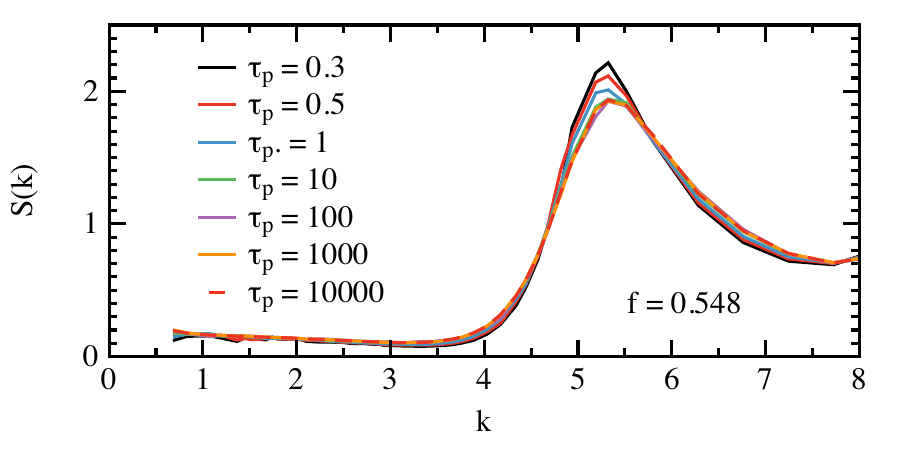}
\caption{\label{sqv0}The persistence time dependence of the steady-state structure factor $S(k)$
for $f=0.548$. The peak height decreases with increasing persistence time and then saturates aroung $\tau_p(f)$. }
\end{figure}

%The results discussed above characterize the motion of the particles in the quiescent or sheared stationary state.
When analyzing the dynamics
in passive systems, one usually tries to make connection between the average distribution of the particles and their dynamics. To check how
the average arrangement of the particles in our active systems changes with increasing persistence time we evaluated the steady-state structure factor
\begin{equation}
S(k) = \frac{1}{N} \left< \sum_{n,m} e^{i \mathbf{k} \cdot (\mathbf{r}_n - \mathbf{r}_m)} \right>.
\end{equation}
In Fig.~\ref{sqv0} we show that the peak height of the structure factor initially decreases with increasing persistence time, which nicely correlates with
relaxation getting faster and viscosity decreasing. The peak height then saturates at persistence time around $\tau_p(f)$. However, the structure factors
for $\tau_p\ge\tau_p(f)$ still look liquid-like \cite{homogeneous}. It is not obvious at all from these structure factors that the MSD exhibits two ballistic regimes
and $F_s(k;t)$ is well fitted by a compressed exponential. We conclude that to describe the dynamics of extremely persistent dense active fluids one cannot rely upon
static structure factors only. 

We presented here a new class of extremely persistent active matter systems. Whereas earlier investigations \cite{Mandal2021,Keta2023}
revealed systems that relax on the time scale of the self-propulsion and exhibit intermittent dynamics with system-size spanning elastic and plastic events,
we uncovered systems that relax on the time scale that, in the large persistence time limit, depends only on the strength of the self-propulsion.
Curiously, the single-particle motion exhibits two ballistic regions separated by a superdiffusive regime. Classic signatures of two-step relaxation
are absent both in the mean square displacement and in the intermediate scattering function.
Many properties that quantify the large persistence time limit of the relaxation depend on the strength of
the active forces as a power law. 

We expect that for higher volume fractions there is a transition between the regime in which the relaxation becomes independent of the persistence
time of the self-propulsion, which is the regime we analyzed, and the regime in which the system flows only on the time scale of the self-propulsion,
which is the regime investigated earlier \cite{Mandal2021,Keta2023}. At a fixed volume fraction the transition would be driven by the strength of the
active forces while at a fixed strength of the active forces it would be driven by the density. We hope that future work will  determine the corresponding
phase diagram, which would be the three-dimensional analog of the diagram uncovered by Liao and Xu \cite{LiaoXu}.

Finally, while for small and moderate persistence times there are approximate theories that can be used to describe the relaxation in active fluids
\cite{SzamelMCT,LiluashviliMCT,FengMCT1,FengMCT2,DebetsMCT}, these theories are not expected to work in the large persistence time limit. Thus, the
discovery of a new different paradigm of extremely persistent active fluids with non-trivial power laws call for additional theoretical work.

We thank L. Berthier and P. Sollich for discussions and comments on the manuscript. Part of this work was done when GS was on sabbatical at
Georg-August Universit\"at G\"ottingen. He thanks his colleagues there for their hospitality.
We gratefully acknowledge the support of NSF Grant No.~CHE 2154241.


\begin{thebibliography}{99}
\bibitem{method}
To approximately quantify $\tau_p(f)$ we fit $v^2$ for small $\tau_p$ to a power law,
which approximately describes the small $\tau_p$ behavior,
and determine when this power law is equal to the average of the large $\tau_p$ value of $b$.
We find that a reasonable value of $\tau_p(f)$ is approximately twice when the power law equals the average value.
\bibitem{homogeneous} The absence of small wavevector peaks implies that the systems are homogenenous. We confirmed this observation
  by evaluating local density histograms at several simulated state points.
\end{thebibliography}
\end{document}